\documentclass[08pt,twocolumn,preprintnumbers,amsmath,amssymb,nofootinbib,prl]{revtex4-1}

\usepackage{amsmath,amssymb,graphicx}
\usepackage{float}
\usepackage{latexsym} 
\usepackage{amsmath,amsthm}
\usepackage{ifpdf}
\usepackage{epstopdf}
\usepackage[dvipsnames]{xcolor}
\usepackage{epsfig}
\usepackage{dcolumn}
\usepackage{bm}
\usepackage{braket}
\usepackage[symbol]{footmisc}
\usepackage{soul}
\usepackage{xcolor}
\usepackage{graphicx}
\usepackage{balance}

\newcommand{\beq}{\begin{equation}}
\newcommand{\eeq}{\end{equation}}
\newcommand{\bea}{\begin{eqnarray*}}
\newcommand{\eea}{\end{eqnarray*}}
\newcommand{\bnea}{\begin{eqnarray}}
\newcommand{\enea}{\end{eqnarray}}

\newcommand{\ba}{\begin{array}{cc}}
\newcommand{\ea}{\end{array}}
\newcommand{\fm}{\end{array} \right]}

\begin{document}
\title{Energy transport in diffusive waveguides}
\author{Kevin J. Mitchell$^1$, Vytautas Gradauskas$^1$, Jack Radford$^1$, Ilya Starshynov$^1$, Samuel Nerenberg$^1$,  Ewan M. Wright$^2$, Daniele Faccio$^{1,2}$}
\email{daniele.faccio@glasgow.ac.uk}
\affiliation{$^1$School of Physics \& Astronomy, University of Glasgow, G12 8QQ Glasgow, UK\\
$^2$Wyant College of Optical Sciences, University of Arizona, Tucson, Arizona 85721, USA}

\begin{abstract}

The guiding and transport of energy, for example of electromagnetic waves  underpins many technologies that have shaped modern society, ranging from long distance optical fibre telecommunications to on-chip optical processors. Traditionally, a mechanism is required that exponentially localises the waves or particles in the confinement region, e.g. total internal reflection at a boundary. We introduce a waveguiding mechanism that relies on a different origin for the exponential confinement and that arises due to the physics of  diffusion. %
We demonstrate this concept using light and show that photon density waves can propagate as a guided mode along a core-structure embedded in a scattering, opaque material, enhancing light transmission by orders of magnitude and along non-trivial, e.g. curved trajectories. This waveguiding mechanism can also occur naturally, for example in the cerebral spinal fluid surrounding the brain, along tendons in the human body and is to be expected in other systems that follow the same physics  e.g. neutron diffusion.
\end{abstract}

\maketitle
{\bf{Introduction.}} 
The scattering of light is ubiquitous and one might argue, is the fundamental mechanism by which we observe light in nature \cite{carminati,Boas}.  
It is the reason the sky is blue and sunsets are red; it is also the reason snow is white and apparently opaque \cite{nature}. In the presence of scattering, the most generic and precise description of propagation is provided by the radiative transport equation. This describes energy transport through a series of scattering processes and therefore also describes other seemingly unrelated regimes such as neutron diffusion \cite{nuclear} that very much like light, also has important imaging applications \cite{neutron,neutron2,neutron3}. \\
In the {\emph{strong diffusive}} regime, the thickness of the medium, $L$, is much larger than the transport mean free path (over which a ray loses all memory of its original direction, equal to the inverse of the reduced scattering coefficient, $\mu^\prime_s$), $L\gg 1/\mu^\prime_s$ and the absorption coefficient is significantly smaller than the scattering  coefficient, $\mu^\prime_s\gg\mu_a$. The radiative transport equation can then be approximated by a diffusion equation that continues to describe diverse phenomena such as heat, neutron and light diffusion. For the case of a pulse of light, this will broaden out whilst still maintaining a pulse-like shape, albeit strongly broadened. This has been referred to as a `photon density wave', hinting at the collective wave-like nature of the  propagation in the medium that can also exhibit behaviour akin to diffraction and interference \cite{Boas}.\\
However, scattering renders materials opaque. This was one of the main challenges that was overcome with the invention of the optical fibre - the development of high purity glass that can transport light over large distances without suffering from the attenuation due to absorption and scattering. \\
Optical fibres and waveguides also rely on a refractive index contrast between an inner core  and an outer cladding region such that the waves undergo total internal reflection and a consequent exponential localisation in space \cite{waveguides1,waveguides2}. 
It is also possible then to take a more heuristic approach 
where the requirement of exponential mode localisation is used as a physical signature for guiding electromagnetic waves.
Returning to the case in point, a notable feature of light  propagation upon entering a scattering medium is that it will be exponentially attenuated and will therefore be largely scattered backwards. 
This simple observation can therefore be used heuristically to raise the question as to whether this exponential decay of the photon density wave can give rise to  confinement or guiding. The observation of exponential decay alone is not sufficient to guarantee a positive answer to this question: it is a necessary condition but boundary effects are important and may modify the final light intensity distribution. There is also no \emph{a priori} guarantee that the same physical mechanism underlying the exponential decay of light in a scattering medium can also support some form of mode guiding.\\
\begin{figure}[t]
    \centering
    \includegraphics[width=8cm]{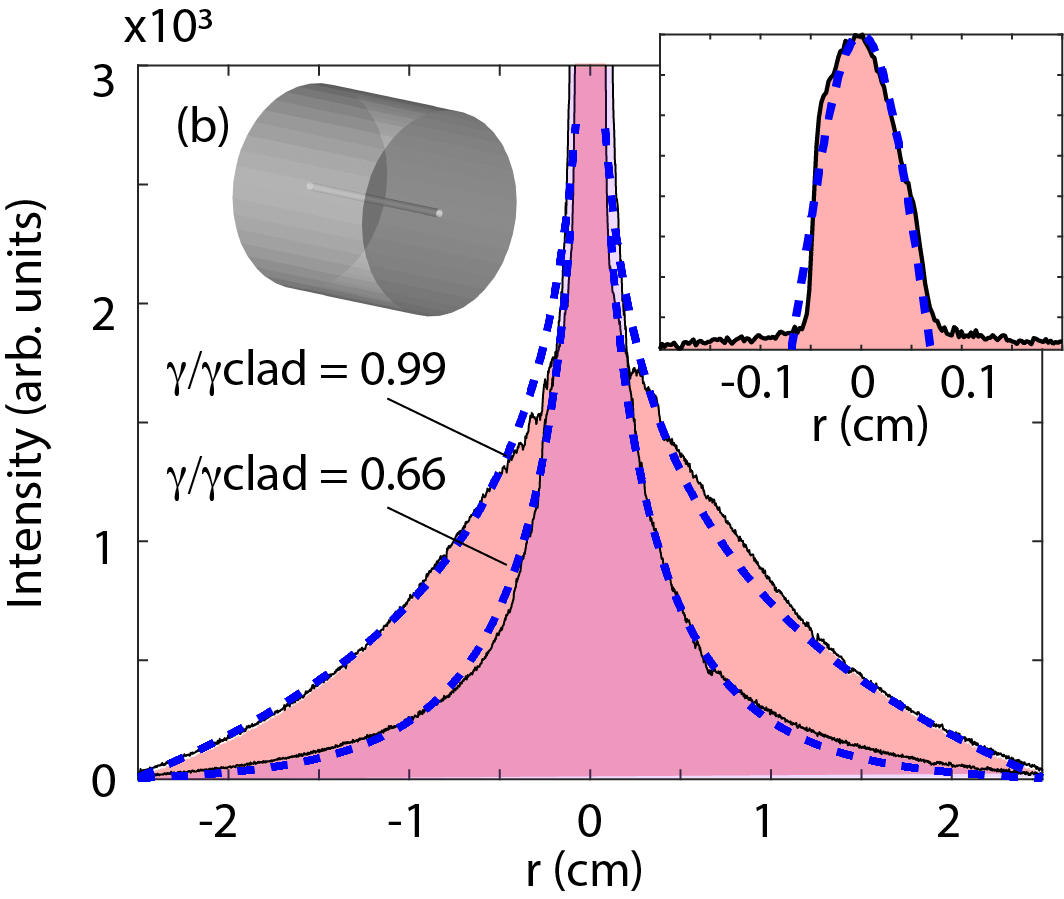}
    \caption{{\bf{Photon density mode guiding, experiments.}} Experimental profiles of light transmitted through a resin cladding structure with a straight 0.5 mm radius core filled with a TiO$_2$:glycerol solution.  Cladding mode profiles are shown for two different TiO$_2$:glycerol solutions (1 mg and 1.5 mg TiO$_2$ in 30 ml glycerol) with fits from the analytical model with Bessel $K_0$ function, where $\gamma/\gamma_{clad}=0.66$ and 0.99, respectively (dashed curves). 
    The inset depicts the profile in the core mode, measured separately due to large dynamic range. This shows a good fit to the predicted Bessel $J_0$ profile that, aside from a vertical scaling factor, was independent from core/cladding details (shown for the 1 mg TiO$_2$ in 30 ml glycerol case). The transmitted energy is measured to be 100x larger compared to a solid (no core) resin cylinder.}
    \label{fig:scatter_core}
\end{figure}
\begin{figure*}[t]
    \centering
    \includegraphics[width=16cm]{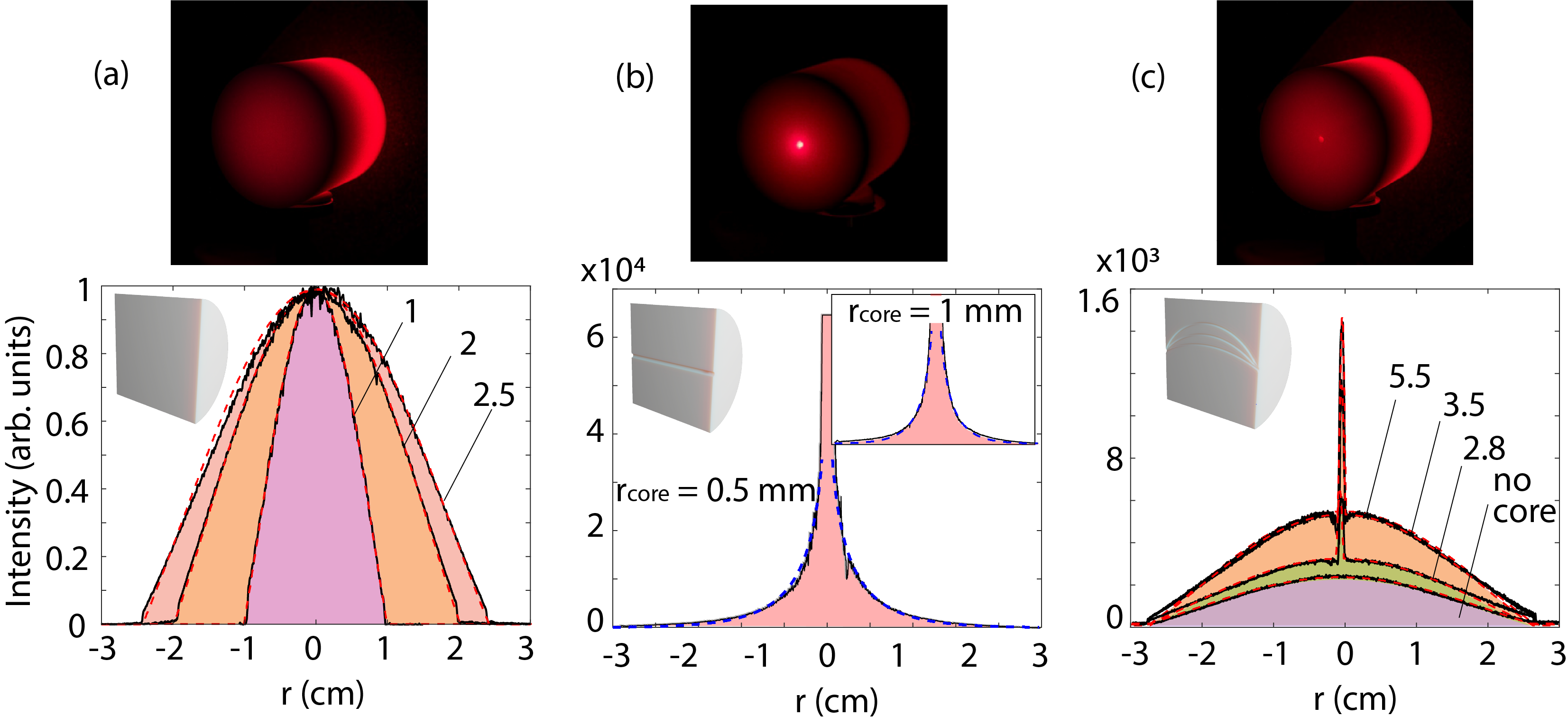}
    \caption{{\bf{Photon density modes, experiments:}} (a) Resin cylinder with no core. The graph shows the measured spatial profiles (horizontal line-outs from full 2D image), for cylinders with radius 1, 2 and 2.8 cm, as indicated next to each curve (the photograph shows an example of 2.8 cm radius cylinder). Solid line - data. Dashed line - parameter-free fits with Bessel functions. (b) Resin cylinder (2.8 cm radius) with a straight (empty) core, Main graph shows experimental line-out for $r_{core}=0.5$ mm (solid curve) and fit with exponential BesselK function (dashed curve). Inset shows same data for $r_{core}=1$ mm.  (c) Resin cylinder (2.8 cm radius) with bent (empty) core.  Three different bend radii are considered, 5.5 cm, 3.5 cm and 2.8 cm, and also a `no core' sample, as indicated in the graph. The 5.5 cm core intensity is $\sim2$x higher than the 3.5 cm core intensity.    
    Top photographs in all figures show the resin structures illuminated with a red laser. Insets in main graphs show a schematic `cut-out' of the resin waveguide structures so as to visualise the internal structure.}
    \label{fig:theory}
\end{figure*}
Here we demonstrate a waveguiding mechanism for photon density waves, whereby a defect
within a uniform strongly scattering medium (e.g. a cylinder of lower diffusion along the propagation direction that we can identify as a ‘core’, analogous to standard optical fibres), effectively guides the energy flow.  The governing equations support the existence of guided modes, supported also by Monte Carlo numerical simulations. Given that the generality of the underlying equations, these results apply also to neutron transport, therefore providing an effective means for the guiding of particles and not only electromagnetic waves.  We perform also experiments that show evidence of `scatter-guiding'. Light is seen to be confined to a broad region that is shaped by the core and is transmitted with more than 2 orders of magnitude more efficiency compared to the case without a core. \\
{\bf{Analytical model for photon density wave modes.}}
 In the limit in which the propagation distance in the scattering medium is $L\gg 1/\mu^\prime_s$ and $\mu^\prime_s\gg\mu_a$ the photon density equation  for the fluence rate $\Phi({\bf r},t)$ is \cite{Boas,Durduran}
\begin{equation}\label{DWE}
 c^{-1} {\partial\Phi\over\partial t} 
 -\nabla D({\bf r}) \nabla \Phi  + \mu_a({\bf r})\Phi  = S({\bf r},t),
\end{equation}
where $S$ is the source, $c$ is the speed of light in the medium and $D\simeq 1/(3\mu_s')$ is the photon diffusion coefficient.  The geometry we consider is a cylindrical  core of radius $R_{core}$ with coefficients $\mu'_{s(core)}$, $\mu_{a(core)}$ that is  surrounded by a coaxial diffusive material cladding of radius $R_{clad}$, with coefficients $\mu'_{s(clad)}$, $\mu_{a(clad)}$ and with air outside.  \\ 
\begin{figure*}
    \centering
    \includegraphics[width=14cm]{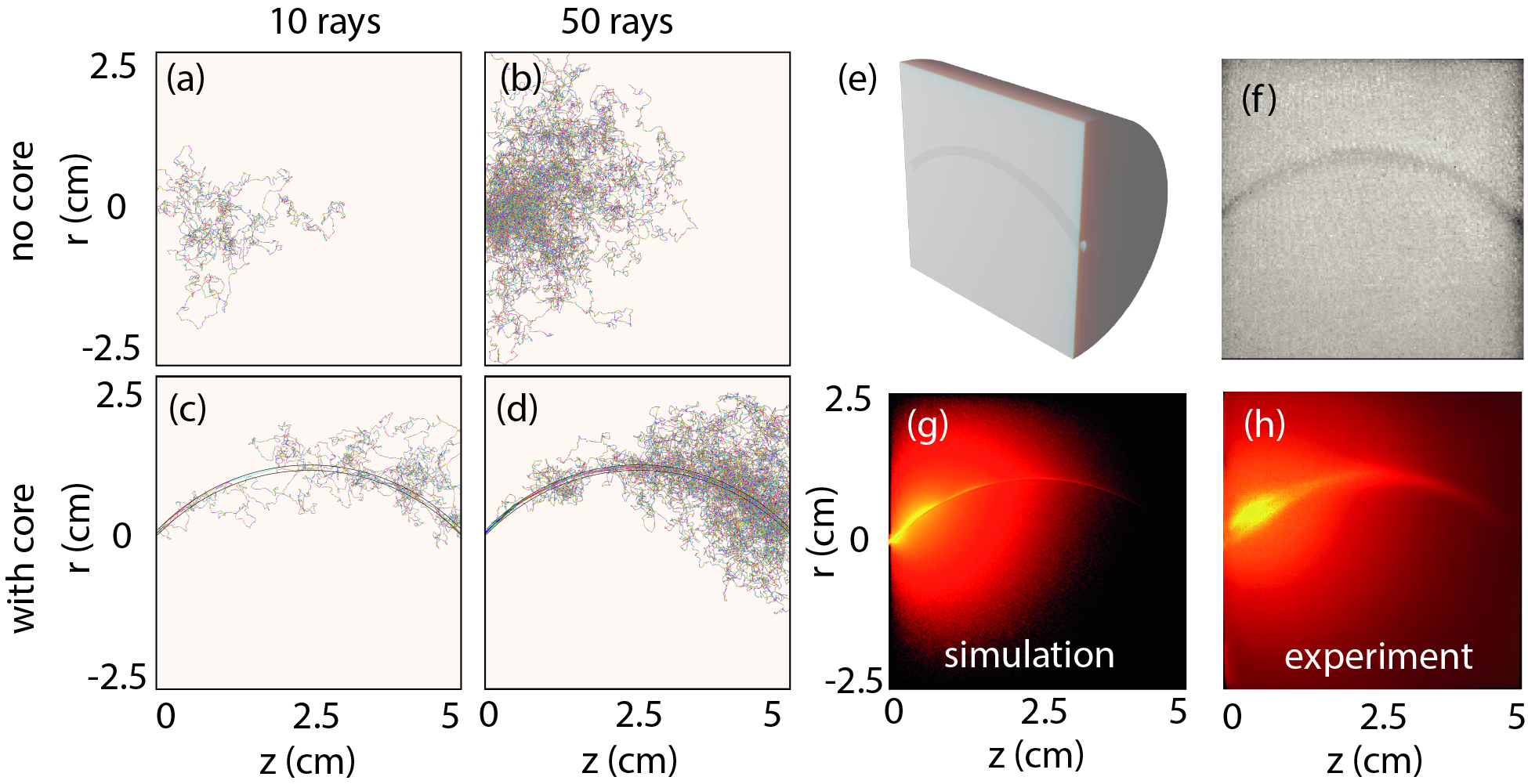}
    \caption{{\bf{Photon density modes in bent waveguides.}} Numerical simulations of photon paths in a solid (no core) resin cylinder (radius 2.8 cm), 10 rays (a) and 50 rays (b) plotted. (c) Same simulations with a 0.5 mm radius, empty core and 10 rays that are detected at the resin structure exit facet (z=5 cm). (d) shows the  same for 50 rays. (e) Schematic of the geometry tested in experiments to observe photon density mode guiding in a bent waveguide in a D-shaped cladding so as to allow visual access to the core and (f), a lateral-view photograph of the 3D-printed waveguide structure. (g) numerical simulation (50 million rays) of the structure in (f). and (h), the experimentally measured light profile.}
    \label{fig:numerics}
\end{figure*}
We remove the source ($S=0$) to look for modal solutions of the form  that decay exponentially $\Phi(r,z) = e^{-\gamma z} \phi(r)$ along the $z$ propagation direction.   The steady-sate photon density equation for modal solutions with cylindrical symmetry  then becomes
\begin{equation}\label{PDEcl}
\nabla_r^2 \phi(r) + (\gamma^2-\gamma_{x}^2) \phi(r) =0
\end{equation}
where $\nabla_r^2$ is the transverse Laplacian, $x$ indicates the core or cladding (i.e. there is an equation each for the core and cladding regions), and $\gamma_x = \sqrt{{\mu_{a(x)}/ D}}\approx \sqrt{3\mu_{a(x)}\mu'_{s(x)}}$ is the extinction rate in the core/cladding material. \\%
Equations~(\ref{PDEcl}) have the form of the well known Helmholtz equations from fibre optics. Here however, the Laplacian term $\nabla_r^2$ describes the effects of photon diffusion as opposed to diffraction. The three key parameters that determine the light propagation regimes as we will see below are $\gamma$ (the modal extinction coefficient), $\gamma_{core}$ and $\gamma_{clad}$ and can be likened  to the propagation constant,  core and cladding refractive index, respectively in the case of traditional optical fibres.  In addition, $\phi(r)$ represents a photon fluence rate that is real and positive, as opposed to the complex electric field envelope in the fibre case. The formal similarity with fibre optics implies that the same type of solutions will apply, and we can exploit that similarity.  In particular, we search for the lowest-extinction mode for our problem that will survive at long distances,  analogous to finding the lowest mode of an optical fibre.  
These modes will preserve their transverse profile with increasing distance, although their overall amplitude will decay exponentially. Finally, we note that a waveguiding structure is obtained by creating a core with $\gamma_{core}<\gamma_{clad}$. This can be achieved by reducing the core scattering coefficient, $\mu'_{s(core)}<\mu'_{s(clad)}$, by reducing the core absorption coefficient, $\mu_{a(core)}<\mu_{a(clad)}$ or by reducing both.\\
{\emph{Mode solutions:}} 
As in the case of standard fibres, we can distinguish between the solution in the core and the solution in the cladding. For all cases where $\gamma<\gamma_{core}$, the core solution is (see SM for full details)
\begin{equation}\label{core}
\phi(r) =  \phi_0 J_0\left ( \sqrt{\gamma^2-\gamma_{core}^2} r \right)  .
\end{equation}
where $\phi_0 $ is a constant and $J_0$ is the zero order Bessel function. As a special case we consider an air core - for this case we take the double limit $D_{core}\rightarrow 0$ and $\mu_{a(core)}\rightarrow 0$, in such a way that $\gamma_{core} = \sqrt{{\mu_{core}/ D_{core}}}\rightarrow 0$.  We then obtain the solution $\phi(r) =  \phi_0 J_0\left (\gamma r \right)$.\\
In the cladding, the solution depends on the relative value of $\gamma$ compared to  $\gamma_{clad}$. For $\gamma<\gamma_{clad}$ we have 
 \begin{equation}\label{Clad_app}
\phi(r) \approx A K_0\left(\sqrt{\gamma_{clad}^2-\gamma^2}~r \right).
\end{equation}
where $A$ is a constant and $K_0$ is the exponentially decaying BesselK function. We therefore have a guided mode in the strict sense, i.e. the mode spatial profile is exponentially localised and indeed has the same functional mode profile as found in standard optical fibres.\\
Some of the results below will involve curved waveguides where, similarly to traditional waveguides, we observe higher modal losses, i.e. $\gamma>\gamma_{clad}$. In this case, the cladding solution will take on the form (see SM)
\begin{equation}
\phi(r) = A  J_0\left(\sqrt{\gamma^2-\gamma_{clad}^2}~r \right),
\end{equation}
i.e. the cladding mode profile switches from a convex exponential decay to a concave Bessel profile.\\
{\bf{Experiments.}} In order to study the main features of photon density waveguiding, we performed experiments in 3D printed resin structures. The resin used in these experiments was characterised using time-of-flight measurements (see e.g. \cite{Swartling:03,Lyons19}) and was measured to have a reduced scattering coefficient of $\mu'_s\simeq35$ cm$^{-1}$ and absorption coefficient of $\mu_a\simeq 0.04$ cm$^{-1}$ (therefore, $\gamma=2$ cm$^{-1}$), in the same range as some biological tissues \cite{Jacques_2013}. Light from a pulsed or CW laser is coupled into the structures via a fibre tip that is placed up against the material or core. A first set of measurements was performed by filling the core with scattering material made of glycerol with absorption coefficient $\mu_a\sim0.06$ cm$^{-1}$  and different concentrations of TiO$_2$ (1-6 mg in 30 ml glycerol) with $\mu'_s$ (that varies linearly with the TiO$_2$ concentration \cite{Swartling:03}) in the range 0.05-0.3 cm$^{-1} $. The resulting core extinction coefficients are therefore $\gamma_{core}=0.09-0.2$ cm$^{-1}$.\\ 
Figure~\ref{fig:scatter_core} shows the measured photon density mode profile at the output (as captured by a CMOS camera from which a line-out is plotted along the horizontal axis, see SM).  The mode profile in the core region is shown in the inset, and was relatively insensitive and only changes (decreases) in amplitude as $\gamma_{core}\rightarrow\gamma_{clad}$. The blue dashed curve shows the expected Bessel profile, Eq.~(\ref{core}).\\
The cladding mode profile is shown for two cases of low TiO$_2$ concentrations ($\gamma_{core}=0.09$ and 0.12). We observe exponentially localised mode profiles that are well fitted (dashed curves) by Eq.~(\ref{Clad_app}) with $\gamma/\gamma_{clad}=0.99$ and 0.66, as indicated in the figure. This demonstrates that it is indeed possible to excite (exponentially localised) photon density modes  in a scattering medium. \\
Figure~\ref{fig:theory}(b) shows the case in which we have an air core with a radius of 0.5 mm or 1 mm (graph inset). Our theory, despite the approximation made in deriving the solutions in this case, correctly predicts an exponentially localised solution that is very well reproduced by the theoretical curve (dashed lines) with $\gamma/\gamma_{clad}=0.44$ and 0.33, respectively.\\
Figure~\ref{fig:theory}(c) shows results with curved air cores for various curvatures (indicated in cm in the figure). In all cases we see that the photon density mode profile in the cladding region has now switched from the convex profile of the exponential decay seen in Fig.~\ref{fig:theory}(b) to the concave, Bessel function shape, as predicted by our theory when $\gamma>\gamma_{clad}$. In analogy with traditional waveguides, the additional losses that increase the value of $\gamma$ can be attributed to the bending of the core.  Despite these higher losses, we note that the transmitted power is significantly larger compared to the light transmitted without a core structure. For the case of the straight scattering core in Fig.\ref{fig:scatter_core} and for the case of the lowest curvature in the air-core waveguides in Fig.~\ref{fig:theory}(c), we found an increase of $\sim110$x and $\sim5$x, respectively (see SM more details and numerical simulations of propagation loss). \\
  In order to gain more insight into what is happening inside the medium, we performed Monte Carlo simulations using the  same structure geometries described above and medium parameters $\mu'_s=15$ 1/cm and $\mu_a= 0.016$ 1/cm (the qualitative features of the results do not depend on these exact values).
Figures~\ref{fig:numerics}(a) and (b) show results for the case of a solid cylinder with no core and in (c), (d) for the case of a curved core, when selecting only 10 rays or 50 rays. In the absence of a core, most of the rays are back-reflected whilst the presence of the core clearly provides a guiding mechanism and the rays accumulate around the curved core and exit the distal end. We note that guiding still takes place despite the photon density modes no longer having a spatial exponential decay profile (compared to the straight core case). Figure~\ref{fig:numerics}(e) shows a schematic view of a structure that consists of a D-shaped cylinder (cladding) with a curved core (bend radius of 28 mm) that is placed 1 mm beneath the flat surface (Fig.~\ref{fig:numerics}(f)). We simulate this experimentally (50 million rays) in Fig.~\ref{fig:numerics}(g) and compare this to an experimental measurement in Fig.~\ref{fig:numerics}(h) (a full set of simulations for various bend curvatures is shown in the SM). The experimental profile agrees  well with the simulated profile and demonstrates that despite the dependence of the photon density mode on the boundary conditions, these modes are robust even to strong modifications of the structure geometry and boundary conditions.\\
\begin{figure}[!t]
    \centering
    \includegraphics[width=7cm]{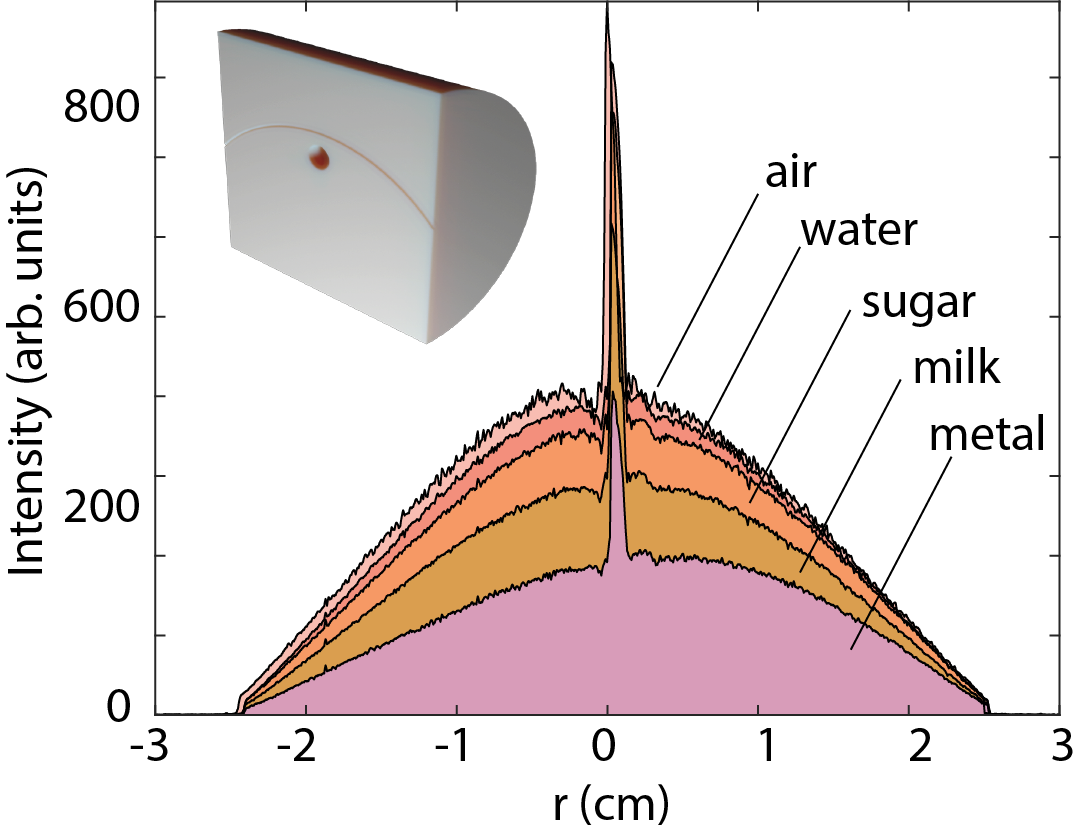}
    \caption{{\bf{Sensitivity to cladding `defects'.}} Experimentally measured output profiles (line-outs from 2D images) with the cylindrical structure shown in the inset (a cut-out is shown of the full cylinder so as to illustrate the internal structure) and with various materials indicated in the graph that fill the `defect' hole.  }
    \label{fig:sensing}
\end{figure}
In Fig.~\ref{fig:sensing} we investigate the effect of a `defect' or inclusion in the cladding. We 3D-printed a structure that has a curved core (curvature radius 25 mm) thus supporting a broad mode,  with a 5 mm radius hole that traverses the whole structure perpendicularly to the plane of the core and is displaced internally from the core by 5 mm (see inset to Fig.~\ref{fig:sensing}). 
The photon density mode maintains its Bessel-like structure but can sense the difference between a hole filled with air or water and it is only when we insert a totally opaque object (metal) that we start to also observe a slight distortion of the mode profile.\\
Interestingly (and partly the inspiration for this work) photon density waveguiding  can also occur naturally. Traditional  optical waveguiding  from a refractive index contrast was similarly first reported in 1842 by Jean-Daniel Colladon to occur naturally in a thin jet of water.  In the case of photon density waves, light propagating through the human head for example is strongly affected by the presence of a relatively transparent cerebral spinal fluid (CSF) that is contained between otherwise dense, scattering layers of bone and gray/white matter. The emphasis in the past has been on the role played by the CSF in limiting light penetration into the brain gray/white matter or as an indicator of neurodegeneration \cite{head1,Wolf1999,Okada2000,Okada2003_CSFmodeling, Dehghani2000,Ancora2018} but could in the future be used as a route to transfer light across larger regions or even across the whole brain, as seen in Monte Carlo simulations (see SM). Similarly, other areas of the human body such as tendons, can also conduct light (see SM for a photograph of a human forearm tendon guiding light over several cms) .\\ 
{\bf{Conclusions.}} By inserting a core structure inside an opaque scattering medium, it is possible to excite exponentially localised modes that survive even in the presence of perturbations such as bending, and which improve light transmission by orders of magnitude. We underline that the light guiding discussed here is fundamentally different from previous light-guiding mechanisms, including the process of guiding light in highly anisotropic, i.e. fibrous scattering media, that has been identified in dentin \cite{dentin}.\\
 Looking at the role of the CSF or other structures in the human body from the perspective of a light-guiding problem might offer new insights into how to control and harness photon density modes to access deep-body locations. It is also possible to clear thin channels using spatially-shaped beams in scattering fluids \cite{Baumgartl:10} or in fog with light filaments from high power lasers with applications for example in free space telecommunications \cite{Kasparian:08,Kasparian:16,Kasparian:18}.  The results presented here would suggest the possibility of a photon density mode that follows the optically cleared channel.\\ 
Finally, we have already underlined that the same equations that govern the propagation of photon density waves apply also to neutron transport, implying that the proposed mechanism provides a mechanism to guide also particles and not just waves.
\\

{\bf{Acknowledgments.}} D.F. is supported by the Royal Academy of Engineering through the Chairs in Emerging Technology programme. The authors acknowledge funding  from the Engineering and Physical Sciences Research Council (EPSRC, UK, Grant No. EP/T00097X/1), the UKRI Frontier Research scheme and ONRG.

\balance
\bibliographystyle{unsrt}
\bibliography{sample}

\begin{thebibliography}{10}

\bibitem{carminati}
R.~Carminati and J.C. Schotland.
\newblock {\em Principles of Scattering and Transport of Light}.
\newblock Cambridge University Press, Cambridge, 2021.

\bibitem{Boas}
D.A. Boas, D.H. Brooks, E.L. Miller, C.A. DiMarzio, M.~Kilmer, R.J. Gaudette, and Quan Zhang.
\newblock Imaging the body with diffuse optical tomography.
\newblock {\em IEEE Signal Processing Magazine}, 18(6):57--75, 2001.

\bibitem{nature}
A.~A. Kokhanovsky.
\newblock {\em Light Scattering Media Optics}.
\newblock Springer, New York, 2004.

\bibitem{nuclear}
Dan Gabriel~Cacuci Ed.
\newblock {\em Handbook of Nuclear Engineering}.
\newblock Springer, New York, 2010.

\bibitem{neutron}
I.S. Anderson, R.L. McGreevy, and H.Z. Bilheux.
\newblock {\em Neutron Imaging and Applications}.
\newblock Spriner, New York, 2009.

\bibitem{neutron2}
J.S. Brenizer.
\newblock A review of significant advances in neutron imaging from conception to the present.
\newblock {\em Physics Procedia}, 43:10--20, 2013.
\newblock The 7th International Topical Meeting on Neutron Radiography (ITMNR-7).

\bibitem{neutron3}
E.~Lehmann, D.~Mannes, A.~Kaestner, and C.~Grünzweig.
\newblock Recent applications of neutron imaging methods.
\newblock {\em Physics Procedia}, 88:5--12, 2017.
\newblock Neutron Imaging for Applications in Industry and Science Proceedings of the 8th International Topical Meeting on Neutron Radiography (ITMNR-8) Beijing, China, September 4-8, 2016.

\bibitem{waveguides1}
A.W. Snyder and J.D. Love.
\newblock {\em Optical Waveguide Theory}.
\newblock Chapman and Hall Ltd., New York, 1983.

\bibitem{waveguides2}
D.~Marcuse.
\newblock {\em Theory of Dielectric Optical Waveguides}.
\newblock Academic Press, Cambridge, Massachusetts, 2012.

\bibitem{Durduran}
Durduran T., Choe R., Baker WB., and Yodh AG.
\newblock Diffuse optics for tissue monitoring and tomography.
\newblock {\em Rep Prog Phys.}, 73(7):076701, 2010.

\bibitem{Swartling:03}
Johannes Swartling, Jan~S. Dam, and Stefan Andersson-Engels.
\newblock Comparison of spatially and temporally resolved diffuse-reflectance measurement systems for determination of biomedical optical properties.
\newblock {\em Appl. Opt.}, 42(22):4612--4620, Aug 2003.

\bibitem{Lyons19}
Ashley Lyons, Francesco Tonolini, Alessandro Boccolini, Audrey Repetti, Robert Henderson, Yves Wiaux, and Daniele Faccio.
\newblock Computational time-of-flight diffuse optical tomography.
\newblock {\em Nature Photonics}, 13:575–579, 2019.

\bibitem{Jacques_2013}
Steven~L Jacques.
\newblock Optical properties of biological tissues: a review.
\newblock {\em Physics in Medicine \& Biology}, 58(11):R37, may 2013.

\bibitem{head1}
Anna Custo, William M~Wells Iii, Alex~H Barnett, Elizabeth M~C Hillman, and David~A Boas.
\newblock Effective scattering coefficient of the cerebral spinal fluid in adult head models for diffuse optical imaging.
\newblock {\em Applied Optics}, 45:4747--4755, 2006.

\bibitem{Wolf1999}
Martin Wolf, Matthias Keel, Vera Dietz, Kurt~Von Siebenthal, Hans~Ulrich Bucher, and Oskar Baenziger.
\newblock The influence of a clear layer on near-infrared spectrophotometry measurements using a liquid neonatal head phantom, 1999.

\bibitem{Okada2000}
Eiji Okada.
\newblock The effect of superficial tissue of the head on spatial sensitivity profiles for near infrared spectroscopy and imaging.
\newblock {\em Opt. Rev.}, 7:375--382, 2000.

\bibitem{Okada2003_CSFmodeling}
Eiji Okada and David~T Delpy.
\newblock Near-infrared light propagation in an adult head model. i. modeling of low-level scattering in the cerebrospinal fluid layer.
\newblock {\em Applied Optics}, 42:2906--2914, 2003.

\bibitem{Dehghani2000}
Hamid Dehghani and David~T Delpy.
\newblock Near-infrared spectroscopy of the adult head: effect of scattering and absorbing obstructions in the cerebrospinal fluid layer on light distribution in the tissue, 2000.

\bibitem{Ancora2018}
Daniele Ancora, Lina Qiu, Giannis Zacharakis, Lorenzo Spinelli, Alessandro Torricelli, and Antonio Pifferi.
\newblock Noninvasive optical estimation of csf thickness for brain-atrophy monitoring.
\newblock {\em Biomedical Optics Express}, 9:4094, 9 2018.

\bibitem{dentin}
Alwin Kienle and Raimund Hibst.
\newblock Light guiding in biological tissue due to scattering.
\newblock {\em Physical Review Letters}, 97, 2006.

\bibitem{Baumgartl:10}
J.~Baumgartl, T.~\v{C}i\v{z}m\'{a}r, M.~Mazilu, V.~C. Chan, A.~E. Carruthers, B.~A. Capron, W.~McNeely, E.~M. Wright, and K.~Dholakia.
\newblock Optical path clearing and enhanced transmission through colloidal suspensions.
\newblock {\em Opt. Express}, 18(16):17130--17140, Aug 2010.

\bibitem{Kasparian:08}
J\'{e}r\^{o}me Kasparian and Jean-Pierre Wolf.
\newblock Physics and applications of atmospheric nonlinear optics and filamentation.
\newblock {\em Opt. Express}, 16(1):466--493, Jan 2008.

\bibitem{Kasparian:16}
Lorena de~la Cruz, Elise Schubert, Denis Mongin, Sandro Klingebiel, Marcel Schultze, Thomas Metzger, Knut Michel, Jérôme Kasparian, and Jean-Pierre Wolf.
\newblock {High repetition rate ultrashort laser cuts a path through fog}.
\newblock {\em Applied Physics Letters}, 109(25):251105, 12 2016.

\bibitem{Kasparian:18}
Guillaume Schimmel, Thomas Produit, Denis Mongin, J\'{e}r\^{o}me Kasparian, and Jean-Pierre Wolf.
\newblock Free space laser telecommunication through fog.
\newblock {\em Optica}, 5(10):1338--1341, Oct 2018.

\end{thebibliography}
\onecolumngrid

\cleardoublepage
\onecolumngrid


\begin{center}
\begin{large}
    \textbf{{Energy transport in diffusive waveguides: Supplementary Material}}

\end{large}
\vspace{0.3cm}
Kevin J. Mitchell$^1$, Vytautas Gradauskas$^1$, Jack Radford$^1$, Ilya 

Starshynov$^1$, Samuel Nerenberg$^1$,  Ewan M. Wright$^2$, Daniele Faccio$^{1,2}$

\textit{
$^1$School of Physics \& Astronomy, University of Glasgow, G12 8QQ Glasgow, UK\\
$^2$Wyant College of Optical Sciences, University of Arizona, Tucson, Arizona 85721, USA}
\end{center}

\vspace{0.5cm}

Supplementary material including the detailed analytical model for photon density modes and additional measurements and figures.

\maketitle
\section{Experimental layout}

\begin{figure*}[ht]
    \centering
    \includegraphics[width=12cm]{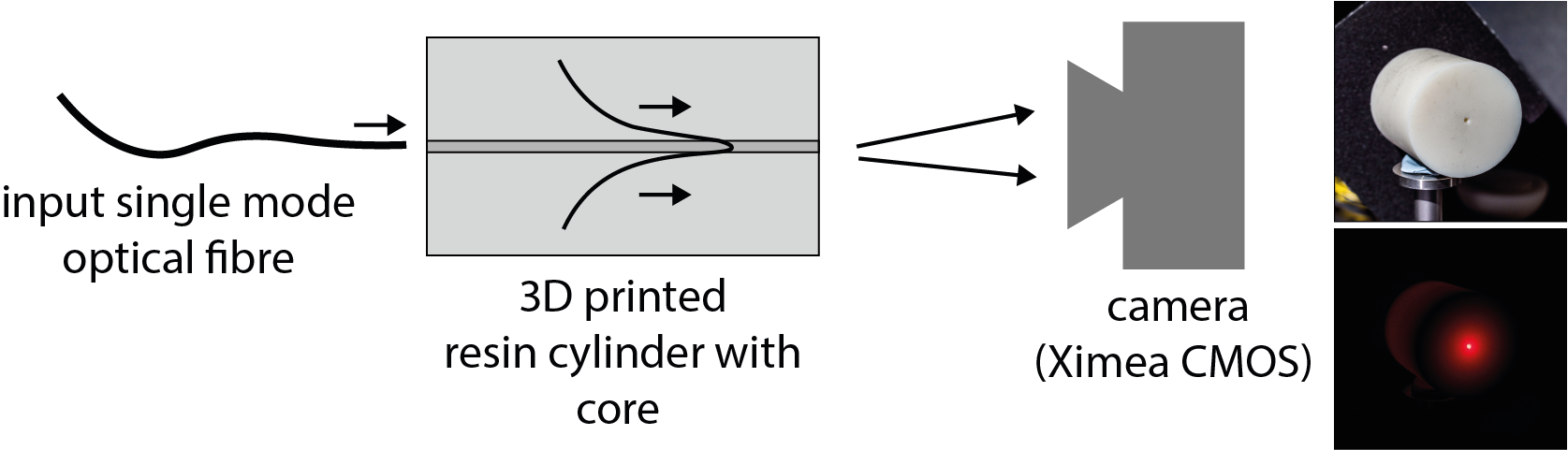}
    \caption{{\bf{Schematic layout of experiments.}}  Light from a laser is coupled into the resin waveguide structure with a single mode fibre. The output photon density mode is imaged onto a CMOS camera. Line-outs of the mode intensity profiles are taken from these 2D images. Insets to the far-right show examples of a resin waveguide and the corresponding visual appearance when illuminated with the laser with the setup shown in the figure. The guided mode is clearly visible at the centre of the resin cylinder.}
    \label{fig:layout}
\end{figure*}
\section{Photon Density Equation}

The photon density equation (PDE) for the fluence $\Phi({\bf r},t)$ is [1,2]
\begin{equation}
 c^{-1} {\partial\Phi\over\partial t} 
 -\nabla D({\bf r}) \nabla \Phi  + \mu_a({\bf r})\Phi  = S({\bf r},t),
\end{equation}
where $S$ is the source, $c$ is the speed of light in the medium, $\mu_a>0$ is a coefficient that relates to the absorption, and $D\simeq 1/(3\mu_s')$ is the photon diffusion coefficient.  The geometry of interest involves a cylindrical core of radius $R_{core}$ surrounded by a coaxial resin cladding of radius $R_{clad}$, with air outside.  To simplify the analysis the following assumptions and approximations are used:
\begin{itemize}

\item[(1)] We assume steady-state conditions so ${\partial\Phi\over \partial t}=0$, and we drop the source $S=0$ to look for modal solutions.

\item[(2)] To simplify we assume that propagation is dominantly along the z-axis and that the solution is cylindrically symmetric 

 
\end{itemize}
With these approximations the steady-state PDE in the core and cladding may be written as
\begin{eqnarray}
-\nabla_r^2 \Phi  - {\partial^2\Phi\over\partial z^2} + \gamma_{core}^2 \Phi &=& 0, \quad 0 < r \le R_{core}   , \nonumber \\
-\nabla_r^2 \Phi  - {\partial^2\Phi\over\partial z^2} + \gamma_{clad}^2 \Phi &=& 0, \quad R_{core} < r \le R_{clad}   ,\end{eqnarray}
where $\nabla_r^2 = \left ( {\partial^2\over\partial r^2} + {1\over r}{\partial\over\partial r}\right )$, and $\gamma_{core} = \sqrt{{\mu_{a(core)}\over D_{core}}}\approx \sqrt{3\mu_{a(core)}\mu'_{s(core)}}$ and $\gamma_{clad} = \sqrt{{\mu_{a(clad)}\over D}}\approx \sqrt{3\mu_{a(clad)}\mu'_{s(clad)}}$ are the bulk spatial decay rates in the core and cladding materials, respectively.

We seek modal solutions of the form
\begin{equation}
\Phi(r,z) = e^{-\gamma z} \phi(r) ,
\end{equation}
where $\gamma$ is the mode spatial decay rate.  The PDEs for the modal solution in the core and cladding then becomes
\begin{eqnarray}\label{PDEcl2}
\nabla_r^2 \phi(r) + (\gamma^2-\gamma_{core}^2) \phi(r) &=& 0, \quad 0 < r \le R_{core}   , \nonumber \\
\nabla_r^2 \phi(r) + (\gamma^2-\gamma_{clad}^2) \phi(r) &=& 0, \quad R_{core} < r \le R_{clad}   ,
\end{eqnarray}
For all cases considered we assume that $\gamma_{clad} > \gamma_{core}$. Key boundary conditions are
\begin{eqnarray}\label{phi_bound}
{\partial\phi\over\partial r} \big |_{r=0}&=& 0 , \nonumber \\
\phi(r = R_{core}-\epsilon) &=&  \phi(r = R_{core}+\epsilon), \nonumber \\D_{core} {\partial\phi\over\partial r} \Big |_{r=R_{core}-\epsilon} &=&  D {\partial\phi\over\partial r} \Big |_{r={R_{core}+\epsilon}} .
\end{eqnarray}
We also generally need $\phi(r=R_{clad})$ and ${\partial\phi\over\partial r} \big |_{r=R_{clad}}$ to assess losses at the cladding air interface.
\vspace{0.3cm}

At this stage we note that Eqs. (\ref{PDEcl2}) have the form of the well known Helmholtz equations from fibre optics. Here however, the Laplacian terms describe the effects of photon diffusion as opposed to diffraction, $\gamma$ is the modal extinction as opposed to the modal propagation constant, and $\gamma_{core}$ and $\gamma_{clad}$ are the bulk extinction of the core and cladding as opposed to the refractive indices.  In addition, $\phi(r)$ represents a photon fluence that is real and positive, as opposed to the complex electric field envelope in the fibre case.  But the formal similarity between our present problem and fibre optics means that the same type of solutions will apply, and here we exploit that similarity.  In particular, we want to find the lowest loss mode for our problem that will survive at long distances, and this is analogous to finding the lowest mode of an optical fibre.  This analogy also verifies that our approach does yield bona fide modes, in that they will preserve their transverse profile with increasing distance although their overall amplitude will decay exponentially.

\section{Core solution}

The equation in the core is
\begin{equation}
\nabla_r^2 \phi(r)= -(\gamma^2-\gamma_{core}^2) \phi(r) = 0, \quad 0 < r \le R_{core}  ,
\end{equation}
where for our conditions $\gamma>\gamma_{core}$.  This equation then has a zero-order Bessel function solution
\begin{equation}\label{core2}
\phi(r) =  \phi_0 J_0\left ( \sqrt{\gamma^2-\gamma_{core}^2} r \right) , \quad 0< r \le R_{core} .
\end{equation}
Hereafter we set the on-axis value $\phi_0=1$ to unity in all cases, and we note that the solution has zero derivative at the origin.

As a special case we want to consider an air core.  For this case we take the double limit $D_{core}\rightarrow 0$ and $\mu_{core}\rightarrow 0$, in such a way that $\gamma_{core} = \sqrt{{\mu_{core}\over D_{core}}}\rightarrow 0$.  We then obtain the solution
\begin{equation}\label{core3}
\phi(r) =  J_0\left (\gamma r \right) , \quad 0< r \le R_{core} .
\end{equation}
For an air core we see that the structure of the solution in the core can be used to extract the modal spatial decay rate $\gamma$.

\section{Cladding solutions}

We now discuss solutions in the cladding for a couple of cases:

\subsection{$\gamma > \gamma_{clad}$}\label{Clad1}

For this case Eq (\ref{PDEcl2}) has Bessel function solutions of the form
\begin{equation}
\phi(r) = A \left [ J_0(\sqrt{\gamma^2-\gamma_{clad}^2}~r) + B Y_0(\sqrt{\gamma^2-\gamma_{clad}^2}~r) \right ],
\end{equation}
with $A$ and $B$ constants. Since the presence of the lossless core tends to lower the modal loss $\gamma$ with respect to the cladding value $\gamma_{clad}$, for this case there must be another source of loss.  This other source of loss must be associated with loss at the cladding outer boundary due to e.g. bending losses (if the core is indeed bent).  In this case we cannot demand that $\phi(r)$ vanishes at the outer-cladding boundary.

\subsection{$\gamma < \gamma_{clad}$}\label{Clad2}

For this case Eq (\ref{PDEcl2}) has modified Bessel function solutions of the form
\begin{equation}
\phi(r) = A \left [ K_0(\sqrt{\gamma_{clad}^2-\gamma^2}~r) + B I_0(\sqrt{\gamma_{clad}^2-\gamma^2}~r) \right ],
\end{equation}
with $A$ and $B$ constants.  This case with $\gamma < \gamma_{clad}$ is relevant in the presence of an unbent core meaning that the spatial decay rate can be less than the cladding value since the core has no losses.  Demanding that the solution vanishes at the cladding boundary yields
\begin{equation}
B = - \frac{K_0(\sqrt{\gamma_{clad}^2-\gamma^2}~R_{clad})}{I_0(\sqrt{\gamma_{clad}^2-\gamma^2}~R_{clad})} .
\end{equation}
giving the cladding solution for $R_{core}< r \le R_{clad}$
\begin{equation}\label{Clad_sol}
\phi(r) = A \left [ K_0(\sqrt{\gamma_{clad}^2-\gamma^2}~r) - K_0(\sqrt{\gamma_{clad}^2-\gamma^2}~R_{clad}) {I_0(\sqrt{\gamma_{clad}^2-\gamma^2}~r)\over I_0(\sqrt{\gamma_{clad}^2-\gamma^2}~R_{clad})} \right ].
\end{equation}
For purposes of comparing with experiment, and similar to fibre optics, we may assume the large cladding limit, which means that we may retain the $K_0(s)$ Bessel term as a reasonable approximation
\begin{equation}\label{Clad_app2}
\phi(r) \approx A K_0(\sqrt{\gamma_{clad}^2-\gamma^2}~r).
\end{equation}
This approximation will be valid as long as $\phi(r)$ in the experiment is close to zero near the cladding-air interface.

There is an interesting limit of Eq. (\ref{Clad_sol}):  For parameters such that $\gamma$ is less than but very close to $\gamma_{clad}$, the argument of the Bessel functions above are small leading to $I_0(s)\approx 1$ and $K_0(s)\approx -\log(s/2)+\gamma_E$.  Using this we find the approximation to the cladding solution
\begin{equation}\label{clad2}
\phi(r) = -A \cdot\log\left ( {r/R_{clad}}\right ) , \quad R_{core}< r \le R_{clad}.
\end{equation}
What is noteworthy about this solution is that it only depends on the cladding radius, not the resin material parameters, or the mode damping $\gamma$.
  
\section{Photon density modes}

In a general setting the photon density modes are found by solving the pair of Helmholtz-like Eqs. (\ref{PDEcl2}) along with the boundary conditions (\ref{phi_bound}) to find the lowest loss mode $\phi(r)$ along with its modal loss rate $\gamma$: This calculation closely follows that for the lowest mode of an optical fibre, with a Bessel function of the first kind and order zero in the core, and a modified Bessel function of the first kind and order zero in the cladding, the loss $\gamma$ being determined by an equation akin to the optical fibre dispersion relation.  Rather than reproduce this analysis here we shall look at some special cases that we have used to compare against the experimental data.

\subsection{Cladding mode with no core}\label{Clad3}

A particular solution of interest is the case without a core, $R_{core}=0$, which has the specific solution with $\gamma>\gamma_{clad}$
\begin{equation}
\phi(r) = J_0(\sqrt{\gamma^2-\gamma_{clad}^2}~r).
\end{equation}
Then demanding that the solution vanishes at the cladding boundary yields
\begin{equation}
\phi(r) = J_0\left ( {j_{0,1}r\over R_{clad}} \right), \quad \gamma = \sqrt{ \gamma_{clad}^2 + \left ({j_{0,1}\over R_{clad}}\right )^2}.
\end{equation}
where $j_{0,1}=2.4048$ is the first root of the Bessel $J_0$ function. Thus we see that the solution for the cladding without the core is a Bessel function, and we find an expression for the spatial decay rate $\gamma>\gamma_{clad}$.

\subsection{Mode for an unbent air core}

In this case the mode loss can be less than the cladding loss, $\gamma<\gamma_{clad}$, meaning that the solutions in cladding and core take the forms in Eqs. (\ref{core3}) and (\ref{Clad_app2}).  Then demanding that $\phi(r)$ be continuous at the core-cladding boundary yields
\begin{eqnarray}
\phi(r) &=& J_0\left ( \gamma r\right) , \quad 0< r \le R_{core} \nonumber \\
&=& J_0\left (\gamma R_{core} \right){K_0(\sqrt{\gamma_{clad}^2-\gamma^2}~r)\over K_0(\sqrt{\gamma_{clad}^2-\gamma^2} R_{core})} , \quad R_{core}< r \le R_{clad}.
\end{eqnarray}
This produces a convex solution akin to what is seen in the experiment.

\subsection{Characterization for a bent air core}

For the case of a bent core there are no strict modes for the system, but approximate solutions can be found  to characterize or fit the spatial profiles at the output.  For the bent core the loss $\gamma$ can exceed the cladding value due to bending losses.  Since $\gamma>\gamma_{clad}$ the solution in the cladding is expected to be of the Bessel function form given above, whereas in the core
\begin{equation}
\phi(r) \approx  A J_0\left ( \gamma r \right) , \quad 0< r \le R_{clad} .
\end{equation}
Then demanding that $\phi(r)$ be continuous at the core-cladding boundary yields
\begin{eqnarray}
\phi(r) &=& J_0\left ( \gamma r \right ) , \quad 0< r \le R_{core} \nonumber \\
&=& J_0\left ( \gamma R_{core}  \right ) {J_0(\sqrt{\gamma^2-\gamma_{clad}^2}~r)\over J_0(\sqrt{\gamma^2-\gamma_{clad}^2} R_{core})}, \quad R_{core}< r \le R_{clad}.
\end{eqnarray}
This produces a concave solution as seen for the bent core.  The approximate loss rate associated with this solution is
\begin{equation}
\gamma = \sqrt{ \gamma_{clad}^2 + \left ({j_{0,1}\over R_{clad}}\right )^2},
\end{equation}
This solution provides an indication as to why the solution changes from convex to concave as the core is bent and/or in general, the modal extinction $\gamma$, increases.

\begin{figure*}[t]
    \centering
    \includegraphics[width=18cm]{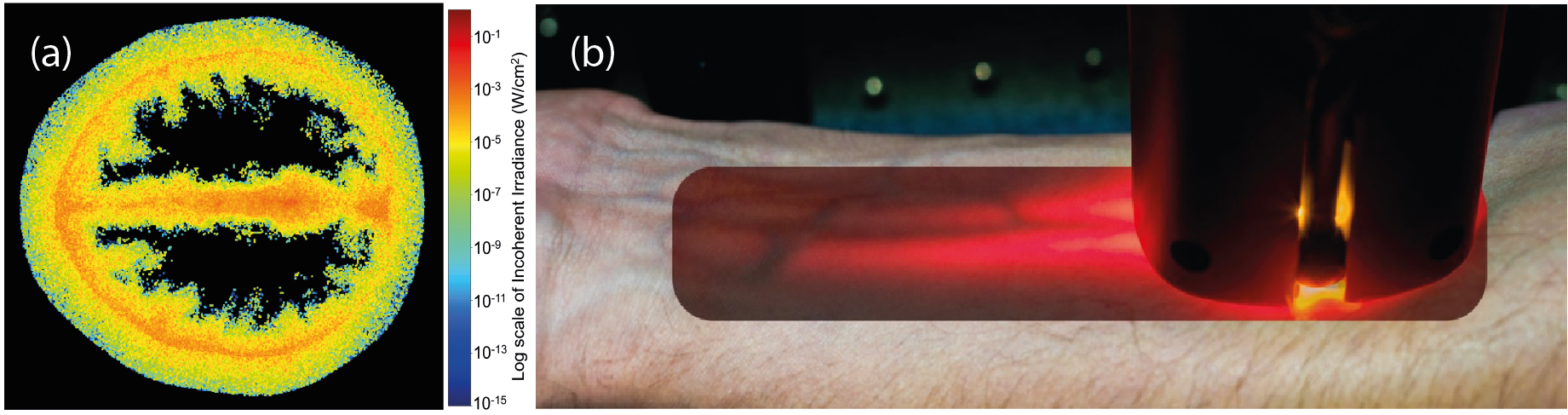}
    \caption{{\bf{Photon density mode guiding examples.}}  (a) Numerical simulation of light transport through the human head, illuminated from the top. Light is predominantly guided by the CSF. (b) Photograph of light guiding in human forearm tendons.}
    \label{fig:nature}
\end{figure*}

\section{Photon density waveguiding in nature.}
Standard optical waveguiding resulting from refractive index contrast was first reported in 1842 by Jean-Daniel Colladon occurring naturally in a thin jet of water. Similarly, photon density waveguiding as a result of diffusion can also occur naturally. In Fig.~\ref{fig:nature} we provide two such examples: (a) shows a Monte Carlo simulation of light propagating through the human head where light propagates as a photon density mode as a result of the layered structure formed by highly diffuse media (the skull and gray/white brain matter) surrounding a weakly diffusive `core' of cerebral spinal fluid (CSF). The importance of the CSF in optical measurements of brain haemodynamics has been investigated previously, with similar numerical results to those shown in Fig.~\ref{fig:nature}(a) albeit with an emphasis on the role played by the CSF in limiting light penetration into the brain gray/white matter or as an indicator of neurodegeneration [3-8].  Figure~\ref{fig:nature}(b) shows a photograph of a human forearm tendon (flexor carpi ulnaris) illuminated from above and showing guiding over several cms as a result of the different scattering and absorption coefficients of the tendon compared to the surrounding muscle tissue. Looking at the role of the CSF or other structures in the human body from the perspective of a light-guiding problem might offer new insights into how to control and harness photon density modes for applications. \\
\vspace{1cm}
\begin{figure*}[t]
    \centering
    \includegraphics[width=12cm]{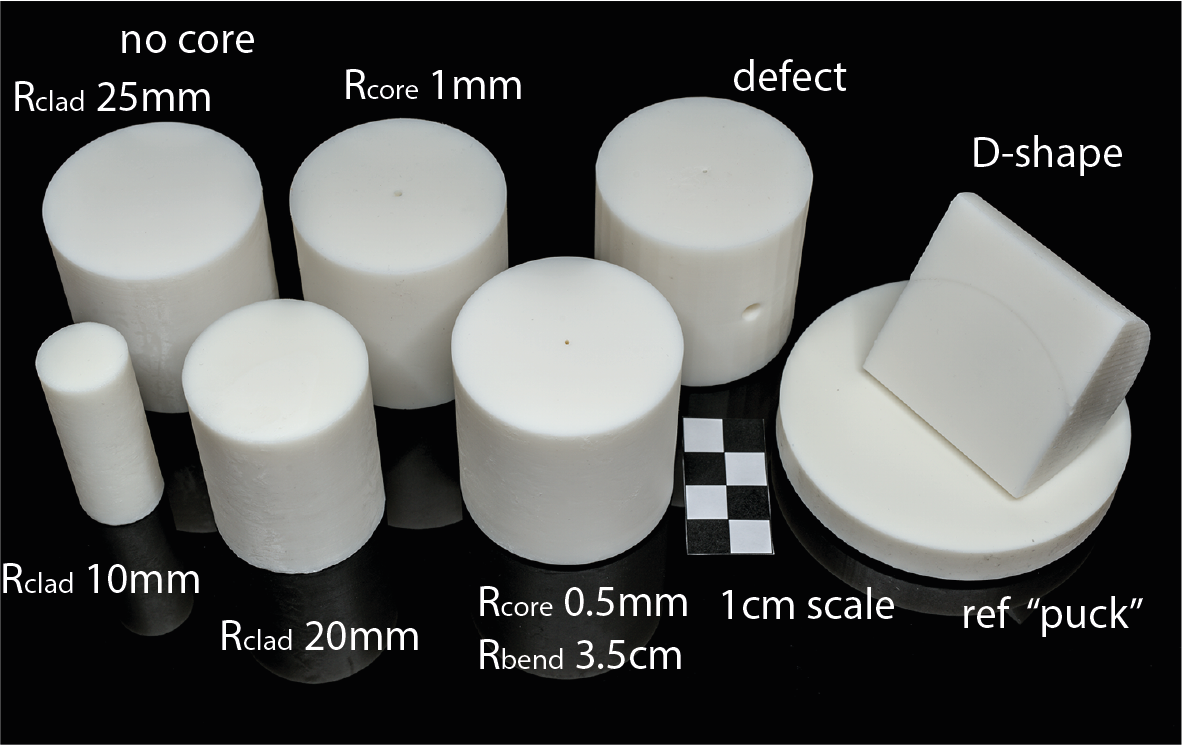}
    \caption{{\bf{Photograph of resin waveguide structures.}}  Photograph of the main resin structures used in this work. Each sample has a label describing the main characteristics such as the outer radius (cladding radius, $R_{clad}$), core radius, bend radius (for the case in which the inner core is bent. The reference `puck' was used to measure resin $\mu'_s$ and $\mu_a$.}
    \label{fig:all_samples}
\end{figure*}

\begin{figure*}[t]
    \centering
    \includegraphics[width=16cm]{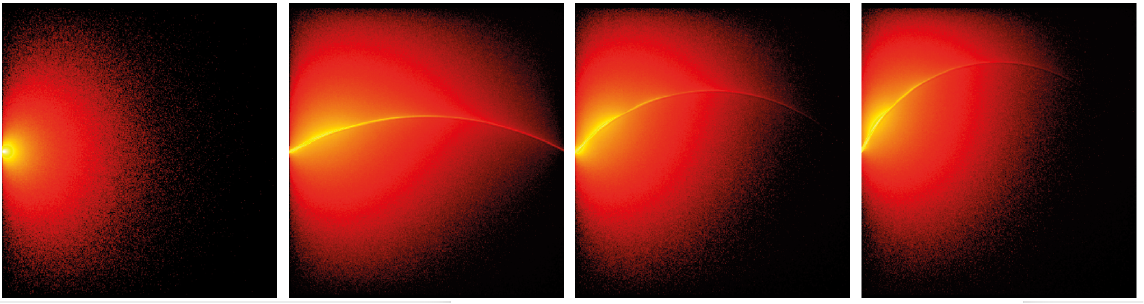}
    \caption{{\bf{Numerical simulations of Photon density modes in bent waveguides.}}  From left to right: resin cylinder with no core; resin cylinder with 1 mm diameter empty core with a bend radius of B1=5.5 cm, B2=3.5 cm and B3=2.8 cm. Data is show in logarithmic scale over 5 decades. All simulations have same parameters as in the main text, Fig.2.}
    \label{fig:Zemax_bent_guides}
\end{figure*}
\begin{figure*}[tb]
    \centering
    \includegraphics[width=14cm]{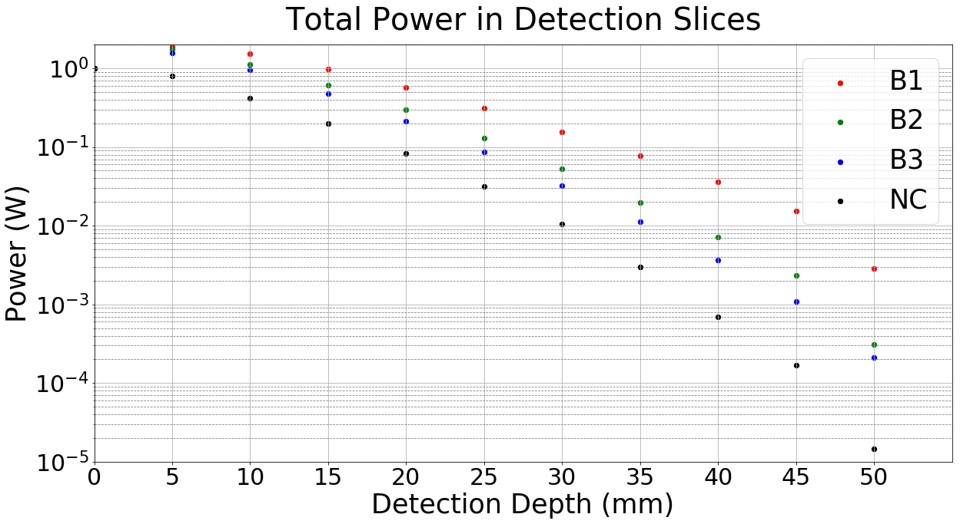}
    \caption{{\bf{Numerical simulations of Photon density mode transmission in bent waveguides.}}  The simulations shown in Fig.~\ref{fig:Zemax_bent_guides} are used to estimate total transmitted power. `NC' refers to the case with no core (no waveguiding). These simulations  show that higher curvatures lead to higher losses. Compared to the `no core' case, waveguiding can transmit up to 100x more energy. The increase in transmitted power is $\sim10$x larger in the simulations compared the experimental value reported in the main text e.g. for the B1 waveguide case. We attribute this difference as a result of additional scattering from surface roughness on the cladding surfaces that result from the 3D printing process (whereas the numerical simulations have perfectly smooth surfaces). }
    \label{fig:Zemax2}
\end{figure*}
\begin{small}

\newpage
\vspace{0.5cm}
\noindent \hangindent=0.55cm [1] D.A. Boas, D.H. Brooks, E.L. Miller, C.A. DiMarzio, M. Kilmer, R.J.Gaudette, and Quan Zhang. Imaging the body with diffuse optical tomography. \textit{IEEE Signal Processing Magazine}, 18(6):57–75, 2001.

\noindent \hangindent=0.55cm [2] Ashley Lyons, Francesco Tonolini, Alessandro Boccolini, Audrey Repetti, Robert Henderson, Yves Wiaux, and Daniele Faccio. Computational time-of-flight diffuse optical tomography. \textit{Nature Photonics}, 13:575–579, 2019.

\noindent \hangindent=0.55cm [3] Anna Custo, William M Wells Iii, Alex H Barnett, Elizabeth M C Hillman, and David A Boas. Effective scattering coefficient of the cerebral spinal fluid in adult head models for diffuse optical imaging. \textit{Applied Optics}, 45:4747–4755, 2006.

\noindent \hangindent=0.55cm [4] Martin Wolf, Matthias Keel, Vera Dietz, Kurt Von Siebenthal, Hans Ulrich Bucher, and Oskar Baenziger. The influence of a clear layer on near-infrared spectrophotometry measurements using a liquid neonatal head phantom, 1999.

\noindent \hangindent=0.55cm [5] Eiji Okada. The effect of superficial tissue of the head on spatial sensitivity profiles for near infrared spectroscopy and imaging. \textit{Opt. Rev.}, 7:375–382, 2000.

\noindent \hangindent=0.55cm [6] Eiji Okada and David T Delpy. Near-infrared light propagation in an adult head model. i. modeling of low-level scattering in the cerebrospinal fluid layer. \textit{Applied Optics}, 42:2906–2914, 2003.

\noindent \hangindent=0.55cm [7] Hamid Dehghani and David T Delpy. Near-infrared spectroscopy of the adult head: effect of scattering and absorbing obstructions in the cerebrospinal fluid layer on light distribution in the tissue, 2000.

\noindent \hangindent=0.55cm [8] Daniele Ancora, Lina Qiu, Giannis Zacharakis, Lorenzo Spinelli, Alessandro Torricelli, and Antonio Pifferi. Noninvasive optical estimation of csf thickness for brain-atrophy monitoring. \textit{Biomedical Optics Express}, 9:4094, 9 2018.

\end{small}

\end{document}